%% file: main.tex
\PassOptionsToPackage{table}{xcolor}
\documentclass[sigconf]{acmart}   

\usepackage{amssymb}
\usepackage{mathtools}
\usepackage{bm}
\usepackage{multirow}
\usepackage{threeparttable}
\usepackage{enumitem}
\usepackage{pifont}
\usepackage{xspace}
\usepackage[ruled,vlined,linesnumbered]{algorithm2e}
\usepackage{balance}   

\definecolor{mygreen}{HTML}{4DAF4A}
\definecolor{myred}{HTML}{E41A1C}
\newcommand{\method}{SparkAlign\xspace}

\newif\ifshowrev
\showrevfalse

\setcopyright{cc}
\setcctype{by}
\copyrightyear{2026}
\acmYear{2026}
\acmConference[CIKM '26]{the 35th ACM International Conference on Information and Knowledge Management}{November 7--11, 2026}{Rome, Italy}
\acmBooktitle{Proceedings of the 35th ACM International Conference on Information and Knowledge Management (CIKM '26), November 7--11, 2026, Rome, Italy}
\acmISBN{979-8-4007-2539-5/2026/11}
\acmDOI{10.1145/3799682.3840078}

\begin{document}

\title{SparkAlign: A Star-Hopping Heuristic for Efficient Unsupervised Plain Graph Alignment}


\author{Boyan Wang}
\authornote{Both authors contributed equally to this research.}
\orcid{0009-0003-3461-7391}
\affiliation{%
  \institution{Hefei University of Technology}
  \city{Hefei}
  \country{China}
}
\email{wangboyan@mail.hfut.edu.cn}

\author{Weijie Feng}
\authornotemark[1]
\authornote{Corresponding author.}
\orcid{0009-0005-1911-4790}
\affiliation{%
  \institution{Hefei University of Technology}
  \city{Hefei}
  \country{China}
}
\email{wjfeng@hfut.edu.cn}

\author{Jinyang Huang}
\orcid{0000-0001-5483-2812}
\affiliation{%
  \institution{Hefei University of Technology}
  \city{Hefei}
  \country{China}
}
\email{hjy@hfut.edu.cn}

\author{Dan Guo}
\orcid{0000-0003-2594-254X}
\affiliation{%
  \institution{Hefei University of Technology}
  \city{Hefei}
  \country{China}
}
\email{guodan@hfut.edu.cn}

\author{Zhi Liu}
\orcid{0000-0003-0537-4522}
\affiliation{%
  \institution{The University of Electro-Communications}
  \city{Tokyo}
  \country{Japan}
}
\email{liu@ieee.org}

\author{Zhuochen Fan}
\orcid{0000-0003-0042-1828}
\affiliation{%
  \institution{Pengcheng Laboratory}
  \city{Shenzhen}
  \country{China}
}
\email{fanzc@pku.edu.cn}

\author{Zhiyong Cheng}
\orcid{0000-0003-1109-5028}
\affiliation{%
  \institution{Hefei University of Technology}
  \city{Hefei}
  \country{China}
}
\email{jason.zy.cheng@gmail.com}

\renewcommand{\shortauthors}{Boyan Wang et al.}

\begin{abstract}
Unsupervised plain graph alignment (UPGA) seeks corresponding nodes between two graphs from topology alone, without attributes or seed anchors.
The dominant embed-then-match paradigm is expensive on both fronts: it repeatedly retrains node embeddings and then solves a costly global assignment, so it scales poorly to large graphs.
We present \method, a training-free heuristic that, inspired by \emph{star-hopping} in celestial navigation, locates a node's counterpart by measuring its heat-diffusion proximity to a shared set of high-confidence \emph{landmark} matches.
Although heuristic in spirit, the descriptor is principled: diffusion makes structurally equivalent nodes comparable \emph{across} graphs, and a short stability analysis justifies a finite diffusion depth.
Matching then uses two sparse, highly parallel strategies that avoid dense global assignment, one favoring speed and one preserving one-to-one consistency on a restricted candidate set.
On three benchmarks \method attains accuracy approaching a structural identifiability reference while running up to two orders of magnitude faster than the strongest baseline and seeding markedly more accurate pseudo-anchors. Our code is available at \url{https://github.com/MaxQ545/SparkAlign}.
\end{abstract}



\maketitle

\input{sections/1-introduction}
\input{sections/2-related-work}
\input{sections/3-method}
\input{sections/4-experiments}
\input{sections/5-conclusion}

\section*{GenAI Usage Disclosure}
Generative AI tools were used during preparation of this manuscript in two
limited capacities: (i) language editing of author-written text for grammar and
concision; and (ii) literature-search assistance to surface candidate related
work, each result of which was subsequently read and verified by the authors
against the primary sources. No part of the methodology, experimental design,
results, or analysis was generated by AI, and the authors take full
responsibility for the content.

\balance
\bibliographystyle{ACM-Reference-Format}
\bibliography{main}

\end{document}

%% file: sections/1-introduction.tex
\section{Introduction}
\label{sec:intro}

Graph alignment identifies corresponding nodes across two graphs and underpins cross-platform user linkage~\cite{wlalign,cone-align,regal,final}, biological network analysis~\cite{galign,walign,grasp}, and knowledge-graph integration~\cite{FGWEA,slotalign,kgalign}.
Most methods lean on side information---node attributes or a seed set of known anchor links~\cite{gradalign+,slotalign,galign,cana}---which in practice is often missing, unreliable, or withheld for privacy~\cite{mmnc,alpine,survey2}.
When neither attributes nor anchors are available the task becomes \emph{unsupervised plain graph alignment} (UPGA): NP-hard, and reliant on topology alone.

State-of-the-art UPGA follows an \emph{embed-then-match} loop: nodes are embedded, a similarity matrix is formed, a matcher commits the most confident pairs, and these pseudo-anchors seed the next round~\cite{mmnc,gradalign,cena}.
Both halves are costly.
\emph{Representation generation} repeatedly retrains or updates embedding models as pseudo-anchors accrue~\cite{mmnc,walign,gradalign}.
\emph{Node matching} faces a quality--scalability tension: direct (row-wise nearest) matching is parallel but produces many-to-one collisions; greedy matching reduces collisions but is sequential; optimal one-to-one assignment is best but costs $O(n^3)$~\cite{mjv,alpine}.
Yet prior methods relieve only one side---training-free representations paired with cubic matching (GRASP), or accelerated matching atop trained embeddings (MMNC)---so none is at once training-free, parallelizable, accelerated, and accurate.

We propose \method, a training-free \emph{heuristic} that attacks both bottlenecks, and whose name reflects its guiding intuition.
In celestial navigation an observer finds a faint target star by \emph{hopping} from a few bright, already-identified stars along recognizable patterns.
\method treats high-confidence matched pairs as such bright ``landmarks'' and locates every remaining node by its heat-diffusion proximity to them---an \emph{anchor-relative} descriptor computed in closed form, with no model to train.
Because diffusion is a deterministic function of topology, structurally equivalent nodes acquire near-identical descriptors even in different graphs (\S\ref{sec:repr})---though not conversely, since structurally similar nodes may also collide---so shared landmarks make the two graphs' representations comparable.
A canonical feature-sorting step supplies permutation-invariant signatures for the first round, when no landmarks exist yet.
For matching we contribute two sparse, fully parallel strategies---a fast collision-reduced matcher and a top-$P$ \emph{sparsified} optimal matcher---that retain one-to-one quality without dense global assignment.

Heuristic though it is, \method rests on a principled footing: we argue its diffusion descriptor is structurally discriminative, and a short stability result answers the natural objection that the diffusion operator is non-contractive, pinning down why a \emph{finite} diffusion depth is optimal (\S\ref{sec:repr}).
Our contributions are:
\begin{sloppypar}
\begin{itemize}[leftmargin=*,topsep=2pt,itemsep=1pt]
  \item a training-free, anchor-relative \emph{heat-diffusion} descriptor for
        UPGA, with a structural-equivalence rationale and a stability result
        that locates the optimal diffusion depth (\S\ref{sec:repr});
  \item two sparse matchers---a parallel fast matcher and a top-$P$ sparsified
        Modified Jonker--Volgenant solver---that cut matching from $O(n^3)$
        toward near-quadratic cost while keeping one-to-one consistency
        (\S\ref{sec:match});
  \item experiments on three benchmarks showing accuracy on par with the
        strongest baseline at up to $315\times$ less
        runtime, with substantially better pseudo-anchor seeding (\S\ref{sec:exp}).
\end{itemize}
\end{sloppypar}

%% file: sections/2-related-work.tex
\section{Related Work}
\label{sec:related}

\noindent\textbf{Graph matching and spectral methods.}
Classical aligners relax the NP-hard quadratic assignment $\min_{\bm P}\lVert\bm P\bm A_s\bm P^{\!\top}-\bm A_t\rVert_F$ over permutation matrices~\cite{isorank,bigalign,grampa}.
Spectral variants match leading eigenvectors of an alignment operator---EigenAlign~\cite{eigenalign}, GRASP~\cite{grasp}---but are noise-sensitive and still need a separate matching step~\cite{survey2}.

\noindent\textbf{Embed-then-match.}
The prevailing paradigm learns node embeddings and then matches them~\cite{survey2}.
Random-walk and GNN encoders---REGAL~\cite{regal}, CONE-Align~\cite{cone-align}, WAlign~\cite{walign}, MMNC~\cite{mmnc}---capture structural proximity but require training and, on plain graphs, a way to make two independently trained spaces comparable.
Optimal-transport aligners cast matching as a (Gromov--)Wasserstein coupling---GWL~\cite{gwlalign}, S-GWL~\cite{s-gwl}, SLOTAlign~\cite{slotalign}, FGWEA~\cite{FGWEA}, CombAlign~\cite{combalign}, FUGAL~\cite{fugal}---which is accurate but leans on iterative transport solvers that are costly at scale.

\noindent\textbf{Iterative / gradual alignment.}
To avoid committing every match at once, gradual methods promote high-confidence pairs to pseudo-anchors and propagate~\cite{cena,gradalign,gtcalign}.
They reuse pseudo-anchors by learning a transform between embedding spaces~\cite{mmnc,walign}, reweighting topology~\cite{deguil,cena,gtcalign}, or recalibrating similarities~\cite{gradalign,final,eigenalign}---at the cost of re-embedding and re-matching each round.

\noindent\textbf{Positioning.}
Many strong baselines mix regimes: GradAlign and CCNE consume seed anchors~\cite{gradalign,ccne}, while CANA and attributed SLOTAlign use node features~\cite{cana,slotalign}; we therefore separate \emph{plain-unsupervised}, \emph{anchor-supervised}, and \emph{attributed} methods in our evaluation (\S\ref{sec:exp}).
\method is plain-unsupervised and uniquely \emph{training-free}: its only ``embedding'' is closed-form diffusion, and it improves by sparsifying---not retraining---the matcher.

%% file: sections/3-method.tex
\section{The \method Method}
\label{sec:method}

\subsection{Problem and Notation}
\label{sec:problem}
Let $\mathcal{G}_s=(\mathcal{V}_s,\mathcal{E}_s,\bm A_s)$ and $\mathcal{G}_t=(\mathcal{V}_t,\mathcal{E}_t,\bm A_t)$ be a source and target graph with $n_s=|\mathcal{V}_s|$ and $n_t=|\mathcal{V}_t|$ unattributed nodes and binary adjacency matrices $\bm A_\star\in\{0,1\}^{n_\star\times n_\star}$; the two sizes may differ.
UPGA seeks a \emph{partial} one-to-one matching $\mathcal{M}=\{(u,v)\}$ between subsets $\mathcal{V}'_s\subseteq\mathcal{V}_s$ and $\mathcal{V}'_t\subseteq\mathcal{V}_t$ (so $n_s\neq n_t$ is allowed), using neither attributes nor \emph{ground-truth} anchors.
We distinguish these from the \emph{pseudo-anchors} $\mathcal{M}^{(r)}$---pairs the algorithm has itself committed by round $r$ and treats as temporary landmarks.
We write $\bm D_\star=\mathrm{diag}(\bm A_\star\bm 1)$ for the degree matrix and $\lVert\cdot\rVert_2$ for the Euclidean norm.

\subsection{Framework Overview}
\begin{figure*}[t]
\centering
\includegraphics[width=0.85\textwidth]{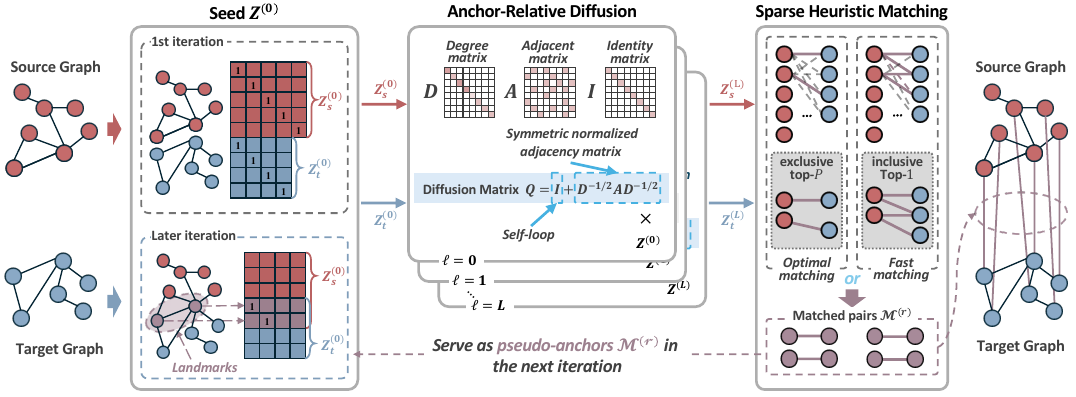}
\Description[Pipeline of one SparkAlign round, from seeding through diffusion to
sparse matching]{A left-to-right pipeline in four stages. At the left, a source
graph and a target graph. In ``Seed $Z^{(0)}$'', the first iteration seeds
one-hot indicator columns for every node, while later iterations seed only at the
committed landmark pairs. In ``Anchor-Relative Diffusion'', the degree matrix,
adjacency matrix and identity matrix combine into the symmetric normalized
diffusion operator $Q=I+D^{-1/2}AD^{-1/2}$, which is applied to the seed for
$\ell=0,\dots,L$ steps to produce the descriptors $Z^{(L)}$ for each graph. In
``Sparse Heuristic Matching'', the descriptors feed two alternative matchers---an
exclusive top-$P$ candidate restriction solved as an optimal assignment, and an
inclusive top-1 fast matcher---each emitting a small set of matched pairs. A
dashed feedback arrow returns those pairs as pseudo-anchors $\mathcal{M}^{(r)}$
to the seeding stage of the next iteration. At the right, the two graphs are
shown with the recovered correspondences drawn between them.}
\caption{\method. Each round embeds unmatched nodes by heat diffusion from
committed landmark pairs (\S\ref{sec:repr}), then commits a few confident
matches with a sparse matcher (\S\ref{sec:match}); commits become landmarks for
the next round.}
\label{fig:archi}
\end{figure*}

\method runs for $R$ rounds; round $r$ performs two stages,
\begin{equation}
\bm H^{(r)}=\underbrace{\mathrm{Embed}\!\left(\mathcal{G}_s,\mathcal{G}_t,\mathcal{M}^{(r-1)}\right)}_{\text{represent}},
\quad
\mathcal{M}^{(r)}=\mathcal{M}^{(r-1)}\cup\underbrace{\mathrm{Match}\!\left(\bm H^{(r)}\right)}_{\text{match}},
\end{equation}
where $\bm H^{(r)}=(\bm H_s^{(r)},\bm H_t^{(r)})$ is the round-$r$ descriptor pair, $\mathrm{Embed}$ embeds the still-unmatched nodes relative to the current pseudo-anchors $\mathcal{M}^{(r-1)}$, and $\mathrm{Match}$ commits a few new high-confidence pairs (Fig.~\ref{fig:archi}).
Committed pairs join the landmark set, so later rounds resolve harder nodes against a richer reference frame.
Here $R$ is the number of rounds, $L$ the diffusion steps, $a_r$ the landmarks at round $r$, $P$ the candidates per node, and $N_{\mathrm{sel}}$ the commits per round.

\subsection{Anchor-Relative Diffusion Representation}
\label{sec:repr}
Given the $a_r=|\mathcal{M}^{(r-1)}|$ pseudo-anchors $\mathcal{M}^{(r-1)}=\{(u_m,v_m)\}_{m=1}^{a_r}$, we seed a signal at each landmark: the columns of $\bm Z_s^{(0)}\in\mathbb{R}^{n_s\times a_r}$ (resp.\ $\bm Z_t^{(0)}$) are one-hot indicators of $u_m$ (resp.\ $v_m$), and unmatched nodes start at zero.
We diffuse this signal over each graph with its own symmetric operator
\begin{equation}
\bm Z_\star^{(\ell)}=\bm Q_\star\,\bm Z_\star^{(\ell-1)},\ \ \ell=1,\dots,L,
\qquad
\bm Q_\star=\bm I+\bm D_\star^{-\frac12}\bm A_\star\,\bm D_\star^{-\frac12},
\label{eq:diffuse}
\end{equation}
for $\star\in\{s,t\}$; the round-$r$ descriptor is the raw diffusion output $\bm H_\star^{(r)}=\bm Z_\star^{(L)}$.
Entry $(i,m)$ of $\bm Z_\star^{(L)}$ measures how strongly heat from landmark $m$ reaches node $i$---node $i$'s coordinates \emph{relative to the landmarks}, exactly the star-hopping intuition.

\noindent\textbf{Why it is discriminative across graphs.}
Heat diffusion is a deterministic function of local topology, so two nodes that play the same structural role receive near-identical diffusion signatures regardless of where they sit~\cite{graphwave,heatkernelpr}; higher-order diffusion further denoises the structural signal against the edge differences between the two graphs~\cite{gdc}.
Anchoring diffusion at a \emph{shared} landmark set turns these per-graph descriptors into a common coordinate system---the diffusion analogue of landmark/pivot embeddings, where representing points by distances to common references yields globally consistent coordinates~\cite{landmarkmds}, and the same principle that lets structural-identity embeddings transfer across disjoint graphs~\cite{regal}.
A true pair $(u,v)$ sees the same landmarks through the same local structure, so their anchor-relative descriptors coincide---which is what makes unsupervised matching possible.
The converse does not hold: the map from local structure to diffusion signature is not injective, so nodes that are merely \emph{similar}---not equivalent---can also receive nearly identical descriptors and become indistinguishable to the matcher.
This is the residual ambiguity the SIR of \S\ref{sec:exp} quantifies, and it bounds what any purely topological descriptor can resolve.
Unlike the fixed structural embeddings of GraphWave and REGAL, our descriptor is re-seeded from \emph{discovered} landmarks each round and thus sharpens as alignment proceeds; unlike landmark/pivot embeddings, its landmarks are discovered rather than supplied.

\noindent\textbf{Cold start.}
In the first round $\mathcal{M}^{(0)}=\emptyset$, so no landmarks exist.
We instead diffuse from the identity, $\bm Z_s^{(0)}=\bm I_{n_s}$ and $\bm Z_t^{(0)}=\bm I_{n_t}$ (each node is its own source), then make the rows comparable across graphs by \emph{canonical sorting}: we sort each row of $\bm Z_\star^{(L)}$ in descending order, keep its top-$d$ entries ($d=\min(n_s,n_t)$, which equalizes the signature length across the two graphs), and $\ell_2$-normalize to form $\bm H_\star^{(r)}$.
Sorting discards the arbitrary node-index ordering, yielding permutation-invariant structural signatures that bootstrap the first set of landmarks.

\noindent\textbf{Stability: why a finite depth is optimal.}
The operator $\bm Q$ in Eq.~\eqref{eq:diffuse} is \emph{not} contractive: $\bm S=\bm D^{-1/2}\bm A\bm D^{-1/2}$ has spectrum in $[-1,1]$ with top eigenvalue $1$, so $\bm Q=\bm I+\bm S$ has eigenvalues in $[0,2]$~\cite{gcn,sgc}.
Iterating $\bm Q$ is therefore a power iteration that amplifies the top eigenvector $\bm u_1\propto\bm D^{1/2}\bm 1$, a purely degree-determined signal.
\begin{lemma}[Over-smoothing limit]
\label{lem:collapse}
For a connected graph let $\bm S$ have eigenpairs $(\lambda_i,\bm u_i)$
ordered $1=\lambda_1>\lambda_2\ge\cdots\ge\lambda_n\ge-1$. For any seed
$\bm x_0$ with $\langle\bm x_0,\bm u_1\rangle>0$ (in particular any nonnegative
seed, since $\bm u_1>0$ entrywise), the $\ell_2$-normalized iterate obeys
\[
\Big\lVert\tfrac{\bm Q^{L}\bm x_0}{\lVert\bm Q^{L}\bm x_0\rVert_2}-\bm u_1\Big\rVert_2
=O\!\Big(\big(\tfrac{1+\lambda_2}{2}\big)^{L}\Big)\xrightarrow{L\to\infty}0 .
\]
\end{lemma}
Each landmark column of $\bm Z_\star^{(L)}$ is such an iterate, so by Lemma~\ref{lem:collapse} every column aligns with $\bm u_1$ and $\bm Z_\star^{(L)}$ approaches a rank-one matrix $\bm u_1\bm c^{\!\top}$: node $i$'s descriptor row degenerates to $u_{1,i}\,\bm c^{\!\top}$---a single degree-determined scalar times a shared vector---so structurally distinct nodes become indistinguishable, the graph analogue of over-smoothing~\cite{oversmoothli,oversmoothoono}.
This degeneration (together with the $\Theta(2^{L})$ growth of the iterate) worsens with depth, whereas too small an $L$ leaves descriptors local and noisy; diffusion depth is therefore a locality--smoothing trade-off with a finite optimum---the peak-then-decay behavior we observe when sweeping $L$ (\S\ref{sec:exp}), and we operate at a small $L$ where magnitudes stay bounded.
We use $\bm Q=\bm I+\bm S$ rather than $\bm S$ because the self-loop keeps the spectrum in $[0,2]$: this removes the sign oscillation a (near-)bipartite $\bm S$ would suffer and, since the power-iteration rate $(1{+}\lambda_2)/2$ exceeds $\lambda_2$, slows this degeneration, widening the useful range of $L$.
A contractive scaling (e.g.\ a lazy walk $\alpha\bm I+(1{-}\alpha)\bm S$) shares $\bm S$'s eigenvectors and the same degeneration, only bounding magnitude and shifting this useful window---which our grid search over $L$ already absorbs---so we keep the simplest operator.

\subsection{Sparse Heuristic Matching}
\label{sec:match}
Both matchers act on the unmatched sets $\mathcal{V}_s^{(r)},\mathcal{V}_t^{(r)}$ (each of size $n_r$) and commit only the $N_{\mathrm{sel}}$ most confident pairs per round; confidence is the descriptor distance, which also induces the ranking used by Hits@1 and MRR.

\noindent\textbf{Fast matching ($\method_f$).}
Each source $u$ proposes its nearest target $v_u^\star=\arg\min_v\lVert\bm h_{s,u}-\bm h_{t,v}\rVert_2$; all proposals are computed in parallel.
We globally sort proposals by distance and accept the best $N_{\mathrm{sel}}$ as new landmarks.
This curbs nearest-neighbor matching's many-to-one collisions---a contested target is claimed by at most one pair per round---with no sequential dependency.

\noindent\textbf{Sparse optimal matching ($\method_o$).}
For strict one-to-one quality we restrict each source $u$ to its $P$ nearest targets $\mathcal{N}_P(u)$ and set every other entry of the cost matrix to $+\infty$,
\begin{equation}
\bm C_{u,v}=
\begin{cases}
\lVert\bm h_{s,u}-\bm h_{t,v}\rVert_2,& v\in\mathcal{N}_P(u),\\
+\infty,&\text{otherwise.}
\end{cases}
\end{equation}
A Modified Jonker--Volgenant solver~\cite{mjv} returns the optimal injective assignment---exact within the candidate set---committing only the $N_{\mathrm{sel}}$ lowest-cost pairs.
The $+\infty$ pairs are never selected, so the solver's augmenting-path search skips almost all of the $n_r^2$ entries; in practice the matcher is dominated by forming the cost matrix, not by a cubic solve.

\noindent\textbf{Complexity.}
Per round, diffusion is $L$ sparse matrix products, $O(L\lvert\mathcal{E}\rvert\delta)$ for $\lvert\mathcal{E}\rvert$ edges and descriptor width $\delta$ (the landmark count $a_r$, or the cold-start dimension $d$); pairwise distances cost $O(n_r^2 \delta)$; fast matching adds $O(n_r^2)$, and the sparse-optimal matcher's cost matrix is likewise $O(n_r^2)$, the candidate restriction ($P\!\ll\! n_r$) keeping its solver clear of a cubic blow-up in practice.
In the \emph{anchor} rounds $\delta=a_r\ll n$, so distances ($O(n_r^2\delta)$) dominate and \method is near-quadratic. Cold start is heavier---$\delta=d=\min(n_s,n_t)$ makes its diffusion, sort, and distances $O(L\lvert\mathcal{E}\rvert n)$, $O(n^2\log n)$, $O(n_r^2 d)$---but cheap on a GPU, where wall-clock follows \emph{work} and serial \emph{depth} jointly: these steps are high-work yet embarrassingly parallel (shallow), the depth floor being the $R$ sequential rounds and the augmenting-path solve. Reported runtimes include it.

%% file: sections/4-experiments.tex
\section{Experiments}
\label{sec:exp}

\begin{sloppypar}
\noindent\textbf{Datasets.}
We use three real-world alignment benchmarks: Facebook--Twitter~\cite{facebook-twitter} ($\sim$1.0K nodes), DBLP1--DBLP2~\cite{dblp} ($\sim$2.2K), and Arxiv1--Arxiv2~\cite{arxiv} ($\sim$18.8K), plus randomly edge-perturbed variants.
\end{sloppypar}

\begin{sloppypar}
\noindent\textbf{Baselines, by regime.}
We separate methods by what they consume.
\emph{Plain-unsupervised} (our setting): FINAL~\cite{final}, REGAL~\cite{regal}, GRASP~\cite{grasp}, CONE-Align~\cite{cone-align}, MMNC~\cite{mmnc}, GWL~\cite{gwlalign}, FUGAL~\cite{fugal}, and Alpine~\cite{alpine}. \emph{Anchor-supervised} (given 10\% seed links, shown for reference only): GradAlign~\cite{gradalign}, WLAlign~\cite{wlalign}, DegUIL~\cite{deguil}. \method$_o$/$_f$ are the optimal and fast variants.
\end{sloppypar}

\noindent\textbf{Metrics.}
We report Hits@1 and MRR; the ranking each metric needs comes from sorting candidates by descriptor distance (\S\ref{sec:match}).
We also report a \emph{structural identifiability reference} (SIR): the fraction of nodes topologically distinguishable from all others (not tied under Weisfeiler--Lehman refinement). It is a \emph{soft} reference, not a hard upper bound, since it credits a match anywhere within an indistinguishable class.
All methods are timed on one machine: a 22-core Intel Xeon Platinum 8470Q, $2{\times}$ NVIDIA RTX PRO 6000 (96\,GB), and 110\,GB RAM.
\method's hyperparameters are few and forgiving: the diffusion depth is best kept at $L\in[5,10]$---below $5$ the descriptors stay too local, beyond $\sim\!10$ diffusion over-smooths (Lemma~\ref{lem:collapse})---and the per-round commit count is robust over $N_{\mathrm{sel}}\in[10,200]$, where a larger $N_{\mathrm{sel}}$ converges in fewer, faster rounds with accuracy stable up to $\sim\!200$ before it degrades.
The exact per-dataset values are released with our code~\cite{sparkalign-code}. \method is deterministic and depends on no random seed; in practice, however, the numerical precision of the hardware still perturbs absolute figures slightly---GPU and CPU runs can differ in the third decimal.

\begin{table}[t]
\centering\footnotesize
\setlength{\tabcolsep}{3.2pt}
\caption{Alignment accuracy (Hits@1\,/\,MRR). \textbf{Bold}/\underline{underline}
$=$ best/2nd among \emph{plain-unsupervised} methods; anchor-supervised methods
(10\% seeds) are shown for reference. SIR is a soft structural-identifiability
reference (\S\ref{sec:exp}), not a hard ceiling. CONE-Align needs equal graph
sizes, so its FB--Tw entries are omitted.}
\label{tab:acc}
\begin{tabular}{l cc cc cc}
\toprule
\multirow{2}{*}{Method} & \multicolumn{2}{c}{FB--Tw} & \multicolumn{2}{c}{DBLP}
  & \multicolumn{2}{c}{Arxiv}\\
\cmidrule(lr){2-3}\cmidrule(lr){4-5}\cmidrule(lr){6-7}
 & H@1 & MRR & H@1 & MRR & H@1 & MRR\\
\midrule
\multicolumn{7}{l}{\textit{Plain-unsupervised}}\\
FINAL  & .1967 & .2763 & .1317 & .2121 & .1853 & .2857\\
REGAL  & .2650 & .3245 & .6992 & .7874 & .6379 & .7203\\
GRASP  & .2280 & .3468 & .0497 & .0983 & .0001 & .0008\\
CONE   & --    & --    & .8168 & .8935 & .5547 & .6667\\
MMNC   & .9140 & .9344 & .8010 & .8858 & .7289 & .8072\\
GWL    & .0000 & .0102 & .0023 & .0073 & .0001 & .0006\\
FUGAL  & .8670 & .8757 & .8703 & \underline{.9237} & .0001 & .0006\\
Alpine & \underline{.9690} & \textbf{.9765} & \underline{.8722} & .9218 & .7778 & .8282\\
\rowcolor{gray!18}$\method_o$ & \textbf{.9700} & \underline{.9723} & \textbf{.8740} & \textbf{.9239} & \textbf{.7803} & \underline{.8347}\\
\rowcolor{gray!18}$\method_f$ & .9540 & .9674 & .8689 & .9215 & \underline{.7779} & \textbf{.8393}\\
\midrule
\multicolumn{7}{l}{\textit{Anchor-supervised (reference; 10\% seeds)}}\\
GradAlign & .9033 & .9359 & .7676 & .8450 & .7367 & .8115\\
WLAlign   & .9141 & .9350 & .8191 & .9001 & .7641 & .7913\\
DegUIL    & .4720 & .5876 & .6188 & .7422 & .5490 & .6497\\
\midrule
\textit{SIR} & .9920 & -- & .8763 & -- & .8076 & --\\
\bottomrule
\end{tabular}
\end{table}

\begin{table}[t]
\centering\footnotesize
\setlength{\tabcolsep}{3pt}
\caption{Efficiency: runtime (s) and peak memory (MB), on the same baselines as
Table~\ref{tab:acc}. \textbf{Bold}/\underline{underline} $=$ best/2nd among
\emph{plain-unsupervised} methods; the \emph{vs.\ Alpine} row gives $\method_f$'s
runtime speedup and $\method_o$'s memory ratio over Alpine (green: better).}
\label{tab:eff}
\begin{tabular}{l rrr @{\hskip 5pt} rrr}
\toprule
\multirow{2}{*}{Method} & \multicolumn{3}{c}{\textit{Runtime (s)}} & \multicolumn{3}{c}{\textit{Memory (MB)}}\\
\cmidrule(lr){2-4}\cmidrule(lr){5-7}
 & FB & DBLP & Arxiv & FB & DBLP & Arxiv\\
\midrule
\multicolumn{7}{l}{\textit{Plain-unsupervised}}\\
FINAL  & 2.6 & 13.1 & 1{,}748 & 961 & 1{,}415 & 50{,}492\\
REGAL  & \underline{1.1} & \underline{2.4} & \underline{102} & \textbf{862} & \textbf{1{,}049} & 18{,}736\\
GRASP  & 1.8 & 4.8 & 1{,}106 & \underline{895} & \underline{1{,}105} & 27{,}801\\
CONE   & -- & 64.0 & 3{,}327 & -- & 1{,}296 & 27{,}761\\
MMNC   & 2.8 & 5.8 & 517 & 931 & 1{,}123 & 28{,}163\\
GWL    & 4.9 & 15.4 & 51{,}132 & 1{,}472 & 1{,}751 & 17{,}707\\
FUGAL  & 133.7 & 358.7 & 13{,}601 & 990 & 1{,}223 & 30{,}884\\
Alpine & 8.5 & 22.2 & 8{,}322 & 1{,}396 & 1{,}623 & 24{,}706\\
\rowcolor{gray!18}$\method_o$ & 3.0 & 7.7 & 1{,}132 & 1{,}450 & 1{,}524 & \textbf{8{,}189}\\
\rowcolor{gray!18}$\method_f$ & \textbf{0.59} & \textbf{0.72} & \textbf{26.4} & 1{,}491 & 1{,}543 & \underline{9{,}711}\\
\textit{vs.\ Alpine} & \cellcolor{mygreen!20}14.4$\times$ & \cellcolor{mygreen!20}30.8$\times$ & \cellcolor{mygreen!20}315.2$\times$ & \cellcolor{myred!20}0.96$\times$ & \cellcolor{mygreen!20}1.07$\times$ & \cellcolor{mygreen!20}3.02$\times$\\
\midrule
\multicolumn{7}{l}{\textit{Anchor-supervised (reference; 10\% seeds)}}\\
GradAlign & 16.0 & 35.3 & 13{,}837 & 1{,}186 & 2{,}199 & 82{,}523\\
WLAlign   & 108.4 & 168.5 & 3{,}968 & 1{,}796 & 2{,}257 & 34{,}013\\
DegUIL    & 9.7 & 11.9 & 478 & 1{,}956 & 2{,}193 & 16{,}831\\
\bottomrule
\end{tabular}
\end{table}

\begin{table}[t]
\centering\footnotesize
\setlength{\tabcolsep}{3pt}
\caption{Left: first-round pseudo-anchor accuracy (Hits@1)---the crux of
iterative UPGA. Right: diffusion-operator ablation (Hits@1); our
$\bm I{+}\bm S$ (self-loop plus symmetric normalization, $\bm S{=}\bm D^{-1/2}\bm A\bm D^{-1/2}$;
$\hat{\bm A}$ is the renormalized variant) wins, as Lemma~\ref{lem:collapse} predicts.}
\label{tab:seed}
\begin{tabular}{l ccc @{\hskip 7pt} l ccc}
\toprule
\multicolumn{4}{c}{\textit{Seed Hits@1}} & \multicolumn{4}{c}{\textit{Operator (Hits@1)}}\\
\cmidrule(lr){1-4}\cmidrule(lr){5-8}
Method & FB & DBLP & Arx & Op. & FB & DBLP & Arx\\
\midrule
REGAL       & .770 & .712 & .655 & $\bm D^{-1}\bm A$ & .8950 & .8675 & .7759\\
MMNC/CENA   & .790 & .841 & .680 & $\bm S$           & .9260 & .8624 & .7727\\
$\method_o$ & \textbf{.980} & \textbf{.885} & \textbf{.770} & $\hat{\bm A}$ & .9320 & .8689 & .7791\\
\rowcolor{gray!18}$\method_f$ & \textbf{.990} & .838 & .760 & $\bm I{+}\bm S$ & \textbf{.9700} & \textbf{.8740} & \textbf{.7803}\\
\bottomrule
\end{tabular}
\end{table}

\subsection{Results}

\begin{sloppypar}
\noindent\textbf{Accuracy approaching the identifiability reference.}
Table~\ref{tab:acc}: $\method_o$ and $\method_f$ match or slightly edge every plain-unsupervised competitor on Hits@1, improving on the strongest baseline (Alpine) by a slim $0.10$--$0.25$ percentage points and approaching the identifiability reference---within $0.23$ percentage points of the SIR on DBLP.
But the decisive advantages lie beyond accuracy: far lower runtime and markedly more accurate pseudo-anchors.
Among the OT baselines, GWL collapses on all three benchmarks (Hits@1$\approx$0), while the stronger FUGAL matches the best methods on the smaller graphs ($.87$ Hits@1 on Facebook--Twitter and DBLP) yet collapses on the 18.8K-node Arxiv pair---underscoring that OT solvers struggle to scale, whereas \method holds $.78$ Hits@1 there at a fraction of the cost.
\end{sloppypar}

\noindent\textbf{Efficiency is the win.}
Table~\ref{tab:eff}: $\method_f$ is fastest on all three datasets, $14$--$315\times$ faster than Alpine, finishing the 18.8K-node Arxiv pair in $26.4$\,s versus Alpine's $2.3$ hours; even the optimal variant $\method_o$ stays competitive, and both variants cut peak memory on Arxiv to $\sim$8--10\,GB ($\sim$3$\times$ less than Alpine).
The training-free descriptor and sparsified matcher remove both efficiency bottlenecks identified in \S\ref{sec:intro}.

\noindent\textbf{Better seeds.}
Iterative UPGA lives or dies by its first pseudo-anchors.
Table~\ref{tab:seed} (left): \method seeds them far more accurately than REGAL or neighborhood-consistency selectors (e.g.\ $.99$ vs.\ $.79$ on Facebook--Twitter), thanks to the permutation-invariant cold-start signatures.
A diffusion-operator ablation (Table~\ref{tab:seed}, right) confirms the design: $\bm I{+}\bm S$ beats the random-walk, symmetric, and renormalized operators on Hits@1 across all datasets.

\begin{figure}[t]
\centering
\includegraphics[width=0.8\linewidth]{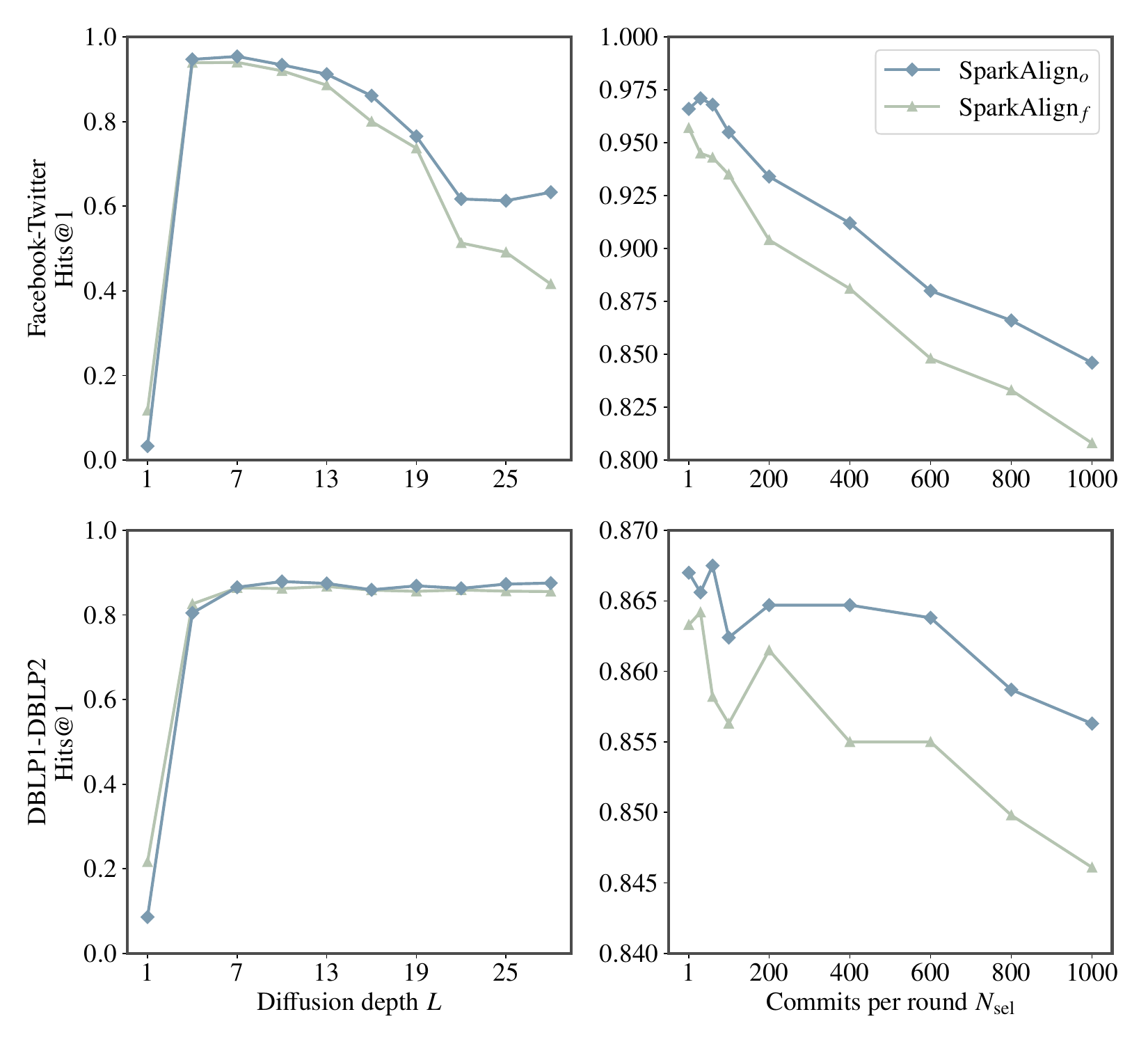}
\Description[Four line charts of Hits@1 against the diffusion depth and the per-round commit count]{A two-by-two grid of line charts, one row per benchmark pair (Facebook--Twitter, DBLP1--DBLP2). The left column plots Hits@1 against the diffusion depth L, the right column against the per-round commit count. Each panel draws one line for $\method_o$ and one for $\method_f$.}
\caption{Effect of the diffusion depth $L$ (left) and the per-round commit count $N_{\mathrm{sel}}$ (right) on Hits@1, for Facebook--Twitter (top) and DBLP (bottom). Note the compressed vertical scales in the right column.}
\label{fig:param}
\end{figure}

\begin{figure}[t]
\centering
\includegraphics[width=\linewidth]{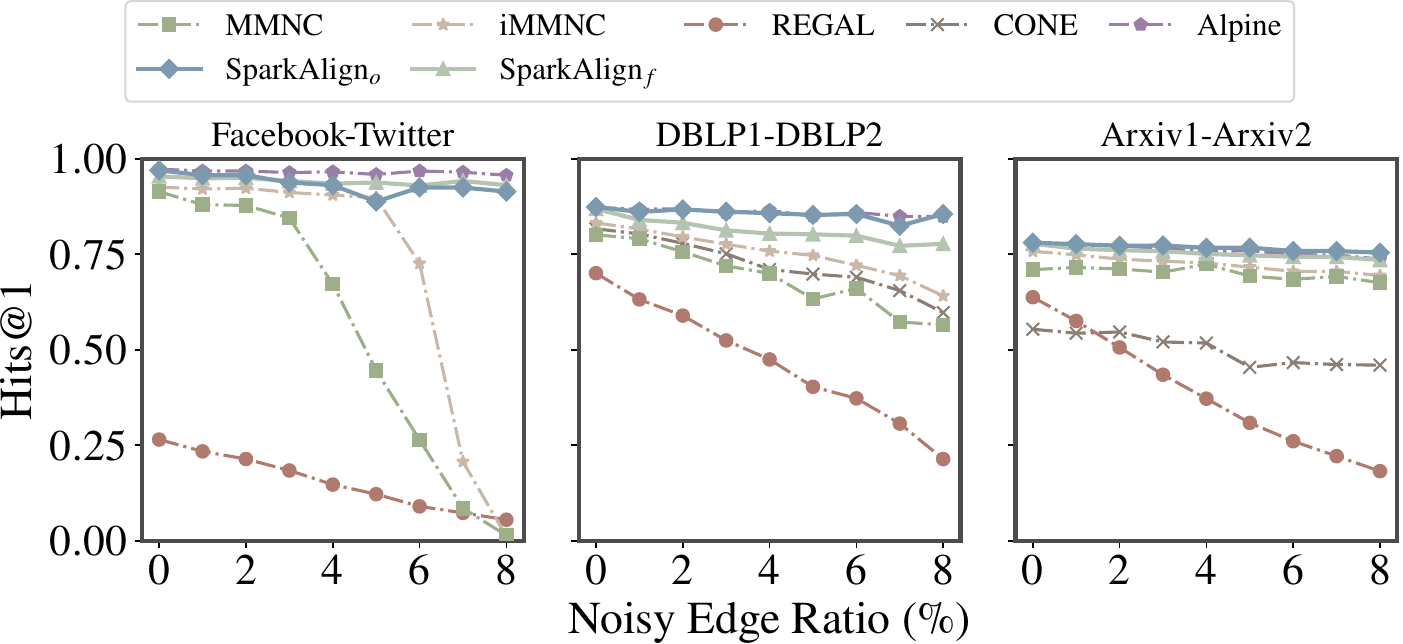}
\Description[Hits@1 versus removed-edge ratio for six methods on three benchmark
graph pairs]{Three side-by-side line charts, one per benchmark pair
(Facebook--Twitter, DBLP1--DBLP2, Arxiv1--Arxiv2). Each plots Hits@1 on the
vertical axis, from 0 to 1, against the percentage of removed edges on the
horizontal axis, from 0 to 8 percent. Six series are shown: MMNC, SparkAlign$_o$,
iMMNC, SparkAlign$_f$, REGAL, CONE and Alpine. All methods decline as noise
increases. On Facebook--Twitter the decline is steepest for every method; on
DBLP1--DBLP2 and Arxiv1--Arxiv2 the two SparkAlign variants decline more slowly
than the baselines and remain close to Alpine across the whole range.}
\caption{Hits@1 vs.\ removed-edge ratio (\%). \method degrades gracefully on the
larger DBLP and Arxiv pairs and stays competitive with Alpine.
}
\label{fig:noise}
\end{figure}

\noindent\textbf{Stability and robustness.}
As Lemma~\ref{lem:collapse} predicts, accuracy rises with $L$, peaks (near $L\!\approx\!7$ on Facebook--Twitter, with a wider plateau on DBLP), then decays as descriptors over-smooth toward the degree signal.
Fig.~\ref{fig:param} sweeps both knobs. The commit count is the milder one: DBLP moves by under one percentage point across $N_{\mathrm{sel}}\!\in\![1,1000]$, whereas Facebook--Twitter peaks near $N_{\mathrm{sel}}\!=\!30$ and then slides as each round commits enough pairs to admit wrong anchors. Both curves are smooth and single-peaked, so a coarse grid is enough to locate a good setting.
Under random edge removal of up to $8\%$ of edges (Fig.~\ref{fig:noise}), \method degrades most gracefully on the larger DBLP and Arxiv pairs---we attribute this to iterative refinement together with its more accurate seeds---while remaining competitive with Alpine.

%% file: sections/5-conclusion.tex
\section{Conclusion}
\label{sec:conclusion}

\begin{sloppypar}
We presented \method, a training-free \emph{star-hopping} heuristic for unsupervised plain graph alignment that attacks both efficiency bottlenecks of the embed-then-match paradigm.
Its anchor-relative heat-diffusion descriptor needs no training yet is principled---structurally discriminative and, by Lemma~\ref{lem:collapse}, stable at a finite, optimal diffusion depth---and its two sparse matchers preserve one-to-one quality without paying for dense $O(n^3)$ assignment.
\method matches state-of-the-art accuracy at up to two orders of magnitude less runtime while seeding markedly better pseudo-anchors.
Extending \method to attributed graphs---folding node attributes into the anchor-relative descriptor---and to the heavier noise of real-world social networks are natural next steps.
\end{sloppypar}

%% file: main.bib
@inproceedings{mmnc,
author = {Tang, Wei and Sun, Haifeng and Wang, Jingyu and Qi, Qi and Wang, Jing and Yang, Hao and Tao, Shimin},
title = {Multi-order Matched Neighborhood Consistent Graph Alignment in a Union Vector Space},
year = {2023},
booktitle = {Proceedings of the International Conference on Research and Development in Information Retrieval (SIGIR)},
pages = {963–972},
}

@inproceedings{facebook-twitter,
author = {Cao, Xuezhi and Yu, Yong},
title = {BASS: A Bootstrapping Approach for Aligning Heterogenous Social Networks},
year = {2016},
booktitle = {European Conference on Machine Learning and Knowledge Discovery in Databases (ECML-PKDD)},
pages = {459–475},
}

@ARTICLE{mjv,
  author={Crouse, David F.},
  journal={IEEE Transactions on Aerospace and Electronic Systems}, 
  title={On implementing 2D rectangular assignment algorithms}, 
  year={2016},
  volume={52},
  number={4},
  pages={1679-1696},
}

@ARTICLE{dblp,
  author={Prado, Adriana and Plantevit, Marc and Robardet, Céline and Boulicaut, Jean-François},
  journal={IEEE Transactions on Knowledge and Data Engineering (TKDE)}, 
  title={Mining Graph Topological Patterns: Finding Covariations among Vertex Descriptors}, 
  year={2013},
  volume={25},
  number={9},
  pages={2090-2104},
}

@misc{arxiv,
    author = {Jure Leskovec and Andrej Krevl},
    title = {{SNAP Datasets}: {Stanford} Large Network Dataset Collection},
    howpublished = {\url{http://snap.stanford.edu/data}},
    month = jun,
    year = 2014
}

@article{gradalign,
author = {Park, Jin-Duk and Tran, Cong and Shin, Won-Yong and Cao, Xin},
title = {On the Power of Gradual Network Alignment Using Dual-Perception Similarities},
year = {2023},
volume = {45},
number = {12},
journal = {IEEE Transaction on Pattern Analysis and Machine Intelligence (PAMI)},
pages = {15292–15307},
}

@inproceedings{survey2,
  title     = {A Survey on Network Alignment: Approaches, Applications and Future Directions},
  author    = {Saxena, Shruti and Chandra, Joydeep},
  booktitle = {Proceedings of the International Joint Conference on Artificial Intelligence, (IJCAI)},
  pages     = {8216--8224},
  year      = {2024},
}

@ARTICLE{wlalign,
  author={Liu, Li and Chen, Penggang and Li, Xin and Cheung, William K. and Zhang, Youmin and Liu, Qun and Wang, Guoyin},
  journal={IEEE Transactions on Knowledge and Data Engineering (TKDE)}, 
  title={WL-Align: Weisfeiler-Lehman Relabeling for Aligning Users Across Networks via Regularized Representation Learning}, 
  year={2024},
  volume={36},
  number={1},
  pages={445-458},
}

@inproceedings{s-gwl,
author = {Xu, Hongteng and Luo, Dixin and Carin, Lawrence},
title = {Scalable {G}romov-{W}asserstein Learning for Graph Partitioning and Matching},
year = {2019},
booktitle = {Advances in Neural Information Processing Systems (NeurIPS)},
}

@INPROCEEDINGS{galign,
  author={Trung, Huynh Thanh and Van Vinh, Tong and Tam, Nguyen Thanh and Yin, Hongzhi and Weidlich, Matthias and Viet Hung, Nguyen Quoc},
  booktitle={IEEE International Conference on Data Engineering (ICDE)}, 
  title={Adaptive Network Alignment with Unsupervised and Multi-order Convolutional Networks}, 
  year={2020},
  pages={85-96},
}

@inproceedings{walign,
author = {Gao, Ji and Huang, Xiao and Li, Jundong},
title = {Unsupervised Graph Alignment with Wasserstein Distance Discriminator},
year = {2021},
booktitle = {Proceedings of the ACM International SIGKDD Conference on Knowledge Discovery and Data Mining (SIGKDD)},
pages = {426–435},
numpages = {10},
}

@INPROCEEDINGS{bigalign,
  author={Koutra, Danai and Tong, Hanghang and Lubensky, David},
  booktitle={ IEEE International Conference on Data Mining (ICDM)}, 
  title={BIG-ALIGN: Fast Bipartite Graph Alignment}, 
  year={2013},
  pages={389-398},
}

@inproceedings{final,
author = {Zhang, Si and Tong, Hanghang},
title = {FINAL: Fast Attributed Network Alignment},
year = {2016},
booktitle = {Proceedings of the ACM SIGKDD International Conference on Knowledge Discovery and Data Mining (SIGKDD)},
pages = {1345–1354},
}

@inproceedings{regal,
  title={{REGAL}: Representation Learning-based Graph Alignment},
  author={Heimann, Mark and Shen, Haoming and Safavi, Tara and Koutra, Danai},
  booktitle={Proceedings of the ACM International Conference on Information and Knowledge Management (CIKM)},
  pages={117--126},
  year={2018}
}

@inproceedings{gradalign+,
author = {Park, Jin-Duk and Tran, Cong and Shin, Won-Yong and Cao, Xin},
title = {GradAlign+: Empowering Gradual Network Alignment Using Attribute Augmentation},
year = {2022},
booktitle = {Proceedings of the ACM International Conference on Information and Knowledge Management (CIKM)},
pages = {4374–4378},
}

@inproceedings{cone-align,
author = {Chen, Xiyuan and Heimann, Mark and Vahedian, Fatemeh and Koutra, Danai},
title = {CONE-Align: Consistent Network Alignment with Proximity-Preserving Node Embedding},
year = {2020},
booktitle = {Proceedings of the ACM International Conference on Information and Knowledge Management (CIKM)},
pages = {1985–1988},
}

@inproceedings{gcn,
  author       = {Thomas N. Kipf and Max Welling},
  title        = {Semi-Supervised Classification with Graph Convolutional Networks},
  booktitle    = {International Conference on Learning Representations (ICLR)},
  year         = {2017},
}

@article{isorank,
author = {Rohit Singh  and Jinbo Xu  and Bonnie Berger },
title = {Global alignment of multiple protein interaction networks with application to functional orthology detection},
journal = {Proceedings of the National Academy of Sciences},
volume = {105},
pages = {12763-12768},
year = {2008},
}

@inproceedings{cena,
  title     = {Joint Link Prediction and Network Alignment via Cross-graph Embedding},
  author    = {Du, Xingbo and Yan, Junchi and Zha, Hongyuan},
  booktitle = {Proceedings of the International Joint Conference on Artificial Intelligence (IJCAI)},
  pages     = {2251--2257},
  year      = {2019},
}

@INPROCEEDINGS{slotalign,
  author={Tang, Jianheng and Zhang, Weiqi and Li, Jiajin and Zhao, Kangfei and Tsung, Fugee and Li, Jia},
  booktitle={2023 IEEE 39th International Conference on Data Engineering (ICDE)}, 
  title={Robust Attributed Graph Alignment via Joint Structure Learning and Optimal Transport}, 
  year={2023},
  volume={},
  number={},
  pages={1638-1651},
}

@inproceedings{FGWEA,
    title = "A Fused {G}romov-{W}asserstein Framework for Unsupervised Knowledge Graph Entity Alignment",
    author = "Tang, Jianheng  and
      Zhao, Kangfei  and
      Li, Jia",
    editor = "Rogers, Anna  and
      Boyd-Graber, Jordan  and
      Okazaki, Naoaki",
    booktitle = "Findings of the Association for Computational Linguistics: ACL 2023",
    month = jul,
    year = "2023",
    address = "Toronto, Canada",
    publisher = "Association for Computational Linguistics",
    pages = "3320--3334",
}

@InProceedings{grampa,
  title = 	 {Spectral Graph Matching and Regularized Quadratic Relaxations: Algorithm and Theory},
  author =       {Fan, Zhou and Mao, Cheng and Wu, Yihong and Xu, Jiaming},
  booktitle = 	 {Proceedings of the 37th International Conference on Machine Learning},
  pages = 	 {2985--2995},
  year = 	 {2020},
  editor = 	 {III, Hal Daumé and Singh, Aarti},
  volume = 	 {119},
  series = 	 {Proceedings of Machine Learning Research},
  month = 	 {13--18 Jul},
  publisher =    {PMLR},
  url = 	 {https://proceedings.mlr.press/v119/fan20a.html},
}

@article{grasp,
  title={{GRASP}: Scalable Graph Alignment by Spectral Corresponding Functions},
  author={Hermanns, Judith and Skitsas, Konstantinos and Tsitsulin, Anton and Munkhoeva, Marina and Kyster, Alexander and Nielsen, Simon and Bronstein, Alexander M and Mottin, Davide and Karras, Panagiotis},
  journal={ACM Transactions on Knowledge Discovery from Data},
  volume={17},
  number={4},
  pages={1--26},
  year={2023},
  publisher={ACM New York, NY}
}

@inproceedings{cana,
  title={CANA: Causal-enhanced Social Network Alignment},
  author={Shao, Jiangli and Wang, Yongqing and Guo, Fangda and Shi, Boshen and Shen, Huawei and Cheng, Xueqi},
  booktitle={Proceedings of the 32nd ACM international conference on information and knowledge management},
  pages={2219--2228},
  year={2023}
}

@article{combalign,
  title={CombAlign: Enhancing Model Expressiveness in Unsupervised Graph Alignment},
  author={Chen, Songyang and Liu, Yu and Zou, Lei and Wang, Zexuan and Lin, Youfang},
  journal={IEEE Transactions on Knowledge and Data Engineering},
  volume={38},
  number={2},
  pages={956--968},
  year={2025},
  publisher={IEEE}
}

@inproceedings{alpine,
  title={Alpine: Partial Unlabeled Graph Alignment},
  author={Petsinis, Petros and Skitsas, Konstantinos and Ranu, Sayan and Mottin, Davide and Karras, Panagiotis},
  booktitle={Proceedings of the 31st ACM SIGKDD Conference on Knowledge Discovery and Data Mining V. 2},
  pages={2315--2325},
  year={2025}
}

@article{gtcalign,
  title={{GTCAlign}: Global Topology Consistency-Based Graph Alignment},
  author={Wang, Chenxu and Jiang, Peijing and Zhang, Xiangliang and Wang, Pinghui and Qin, Tao and Guan, Xiaohong},
  journal={IEEE Transactions on Knowledge and Data Engineering},
  volume={36},
  number={5},
  pages={2009--2025},
  year={2023},
  publisher={IEEE}
}

@inproceedings{eigenalign,
  title={Low rank spectral network alignment},
  author={Nassar, Huda and Veldt, Nate and Mohammadi, Shahin and Grama, Ananth and Gleich, David F},
  booktitle={Proceedings of the 2018 World Wide Web Conference},
  pages={619--628},
  year={2018}
}

@inproceedings{gwlalign,
  title={{G}romov-{W}asserstein Learning for Graph Matching and Node Embedding},
  author={Xu, Hongteng and Luo, Dixin and Zha, Hongyuan and Duke, Lawrence Carin},
  booktitle={International conference on machine learning},
  pages={6932--6941},
  year={2019},
  organization={PMLR}
}

@article{ccne,
  title={Collaborative cross-network embedding framework for network alignment},
  author={Zhang, Hai-Feng and Ren, Guojing and Ding, Xiao and Zhou, Li and Zhang, Xingyi},
  journal={IEEE Transactions on Network Science and Engineering},
  volume={11},
  number={3},
  pages={2989--3001},
  year={2024},
  publisher={IEEE}
}

@article{fugal,
  title={{FUGAL}: Feature-fortified Unrestricted Graph Alignment},
  author={Bommakanti, Aditya and Vonteri, Harshith R and Skitsas, Konstantinos and Ranu, Sayan and Mottin, Davide and Karras, Panagiotis},
  journal={Advances in Neural Information Processing Systems},
  volume={37},
  pages={19523--19546},
  year={2024}
}

@inproceedings{deguil,
  title={DegUIL: Degree-aware graph neural networks for long-tailed user identity linkage},
  author={Long, Meixiu and Chen, Siyuan and Du, Xin and Wang, Jiahai},
  booktitle={Joint European Conference on Machine Learning and Knowledge Discovery in Databases},
  pages={122--138},
  year={2023},
  organization={Springer}
}

@inproceedings{kgalign,
  title={Unsupervised deep cross-language entity alignment},
  author={Jiang, Chuanyu and Qian, Yiming and Chen, Lijun and Gu, Yang and Xie, Xia},
  booktitle={Joint European conference on machine learning and knowledge discovery in databases},
  pages={3--19},
  year={2023},
  organization={Springer}
}

@inproceedings{graphwave,
  title={Learning structural node embeddings via diffusion wavelets},
  author={Donnat, Claire and Zitnik, Marinka and Hallac, David and Leskovec, Jure},
  booktitle={Proceedings of the 24th ACM SIGKDD international conference on knowledge discovery \& data mining},
  pages={1320--1329},
  year={2018}
}

@article{heatkernelpr,
  title={The heat kernel as the pagerank of a graph},
  author={Chung, Fan},
  journal={Proceedings of the National Academy of Sciences},
  volume={104},
  number={50},
  pages={19735--19740},
  year={2007},
  publisher={National Academy of Sciences}
}

@inproceedings{gdc,
  title={Diffusion improves graph learning},
  author={Gasteiger, Johannes and Wei{\ss}enberger, Stefan and G{\"u}nnemann, Stephan},
  journal={Advances in neural information processing systems},
  volume={32},
  year={2019}
}

@techreport{landmarkmds,
  title={Sparse multidimensional scaling using landmark points},
  author={De Silva, Vin and Tenenbaum, Joshua B},
  year={2004},
  institution={technical report, Stanford University}
}

@inproceedings{sgc,
  title={Simplifying graph convolutional networks},
  author={Wu, Felix and Souza, Amauri and Zhang, Tianyi and Fifty, Christopher and Yu, Tao and Weinberger, Kilian},
  booktitle={International conference on machine learning},
  pages={6861--6871},
  year={2019},
  organization={Pmlr}
}

@inproceedings{oversmoothli,
  title={Deeper insights into graph convolutional networks for semi-supervised learning},
  author={Li, Qimai and Han, Zhichao and Wu, Xiao-Ming},
  booktitle={Proceedings of the AAAI conference on artificial intelligence},
  volume={32},
  number={1},
  year={2018}
}

@inproceedings{oversmoothoono,
  title={Graph neural networks exponentially lose expressive power for node classification},
  author={Oono, Kenta and Suzuki, Taiji},
  journal={arXiv preprint arXiv:1905.10947},
  year={2019}
}

@misc{sparkalign-code,
  key          = {SparkAlign},
  author       = {Wang, Boyan},
  title        = {{SparkAlign}: Source Code and Configurations},
  howpublished = {\url{https://github.com/MaxQ545/SparkAlign}},
  year         = {2026}
}
